\documentclass[]{pasj01}
\Received{2021/08/23}%{yyyy/mm/dd}
\Accepted{2021/12/18}%{yyyy/mm/dd}
\usepackage{hyphenat} % allow break
\usepackage[modulo]{lineno}
\usepackage{url}
\usepackage{comment}
\usepackage{upgreek}
    \makeatletter
\def\@fnsymbol#1{\ensuremath{\ifcase#1\or \dagger\or \ddagger\or
   \mathsection\or \mathparagraph\or \|\or **\or \dagger\dagger
   \or \ddagger\ddagger \else\@ctrerr\fi}}
    \makeatother
\begin{document} 

\title{ 
%\LETTERLABEL %%% <-- uncomment for LETTER article  
%\REVIEWLABEL %%% <-- uncomment for REVIEW article  
H$\alpha$ emission in the outskirts of galaxies at z=0.4}

%%% begin:list of authors
% Do NOT capitalize all letters in "textsc".
\author{Rhythm \textsc{Shimakawa}\altaffilmark{1}$^\ast$\thanks{NAOJ Fellow}}
\email{rhythm.shimakawa@nao.ac.jp (RS)}

\author{Masayuki \textsc{Tanaka}\altaffilmark{1}}
\author{Satoshi \textsc{Kikuta}\altaffilmark{2}}
\author{Masao \textsc{Hayashi}\altaffilmark{1}}

\altaffiltext{1}{National Astronomical Observatory of Japan (NAOJ), National Institutes of Natural Sciences, Osawa, Mitaka, Tokyo 181-8588, Japan}
\altaffiltext{2}{Center for Computational Sciences, University of Tsukuba, 1-1-1 Tennodai, Tsukuba, Ibaraki 305-8577, Japan}

%%% end:list of authors

%% `\KeyWords{}' always has to be placed before ``\maketitle'' 
%%  List of Key Words:  https://academic.oup.com/pasj/pages/Pasj_Keywords 

\maketitle

\begin{abstract}
This paper reports detections of H$\alpha$ emission and stellar continuum out to approximately 30 physical kpc, and H$\alpha$ directionality in the outskirts of H$\alpha$-emitting galaxies (H$\alpha$ emitters) at $z=0.4$. 
This research adopts narrow-band selected H$\alpha$ emitters at $z=0.4$ from the emission-line object catalog by \citet{Hayashi2020}, which is based on data in the Deep and Ultradeep layers of the Hyper Suprime-Cam Subaru Strategic Program. 
Deep narrow- and broad-band images of 8625 H$\alpha$ emitters across 16.8 deg$^2$ enable us to construct deep composite emission-line and continuum images. 
The stacked images show diffuse H$\alpha$ emission (down to $\sim5\times10^{-20}$ erg~s$^{-1}$cm$^{-2}$arcsec$^{-2}$) and stellar continuum (down to $\sim5\times10^{-22}$ erg~s$^{-1}$cm$^{-2}$\AA$^{-1}$arcsec$^{-2}$), extending beyond 10 kpc at stellar masses $>10^9$ M$_\odot$, parts of which may originate from stellar halos. 
Those radial profiles are broadly consistent with each other. 
In addition, we obtain a dependence of the H$\alpha$ emission on the position angle because relatively higher H$\alpha$ equivalent width has been detected along the minor-axis towards galaxy disks. 
While the H$\alpha$ directionality could be attributed to biconical outflows, further research with hydrodynamic simulations is highly demanded to pin down the exact cause. 
\end{abstract}

%\linenumbers

%%%%%%%%%%%%%%%%% INTRO %%%%%%%%%%%%%%%%%%
\section{Introduction}\label{s1}

Galaxy outskirts record mass-assembly histories of merger and accretion events in a hierarchical universe \citep{Searle1978,White1978}, and also feedback processes taking place therein \citep{Dekel1986,Heckman1990}. 
Past deep wide-field imaging uncovered various diffuse stellar halos of nearby galaxies and hence diverse buildup histories of outer disks \citep{Courteau2011,Deason2014,Merritt2016,Harmsen2017}, which correlate with their physical properties \citep{Elias2018}. 
Combined analyses with simulations in the cosmological framework of Lambda Cold Dark Matter claim that, in general, high fractions of inner and outer stellar halos are formed by in-situ and ex-situ stellar components, respectively \citep{ZhangJ2018,Merritt2020,Font2020}. 
Furthermore, the diffuse ionized gas and its physical states in the outskirt of nearby galaxies have been well investigated for understanding feedback mechanisms (e.g., \cite{Veilleux2005} and references therein). 
\citet{Zhang2016,Zhang2018} successfully detected H$\alpha+$[N{\sc ii}] emission at $z=$ 0.05--0.2 out to 100 kpc by combining millions of fiber spectra from the Sloan Digital Sky Survey, providing a unique insight into cool gas from associated halos. 
However, it remains significantly challenging to observe the galaxy outskirts beyond the local universe due to cosmological surface brightness dimming, which has hampered direct observations to date from the earlier phase of the mass assembling. 

Nowadays, a giant wide-field imager on the Subaru Telescope with an 8.2 m primary mirror, Hyper Suprime-Cam \citep{Miyazaki2018,Furusawa2018,Kawanomoto2018,Komiyama2018} has enabled very deep wide-field observations under seeing conditions of $<1$ arcsec that provides new insights into the outskirts of nearby galaxies \citep{Fukushima2019,Okamoto2019}. 
In addition, such high-quality imaging data over $>100$ deg$^2$ taken by Hyper Suprime-Cam Subaru Strategic Program (HSC-SSP; \cite{Aihara2018}) can establish even deeper images by stacking numerous objects. 
This enables us to investigate the outskirts of massive galaxies beyond the local universe \citep{Huang2018a,Wang2019}, and their environmental dependence \citep{Huang2018b}. 

With such scientific backgrounds, this research aims at investigating the outskirts of H$\alpha$ line emission from star-forming galaxies at $z=0.4$. 
H$\alpha$ line is used extensively as a tracer of star formation \citep{Kennicutt1998}, which helps us study galaxy outskirts from a unique standpoint. 
Based on the emission-line object catalog available from the Second Public Data Release of HSC-SSP (HSC-SSP PDR2; \cite{Aihara2019,Hayashi2020}), we perform a stacking analysis of H$\alpha$ line and continuum images for 8625 H$\alpha$ emitting galaxies termed as H$\alpha$ emitters at $z=0.4$. 
A wide and deep narrow-band survey by HSC-SSP successfully detects H$\alpha$ emitters at $z=$0.393--0.404 down to $\sim1\times10^{-17}$ erg~s$^{-1}$cm$^{-2}$ over 16.8 deg$^2$ (section~\ref{s2}). 
We investigate radial profiles of their composite H$\alpha$ and continuum images in different stellar mass bins, and study star formation in the outskirts of H$\alpha$ emitters (section~\ref{s3}). 
We also discuss other possible contributions to H$\alpha$ emission on the galaxy outskirts by shape-aligned stacked images, and lastly summarize the entire flow of this work (section~\ref{s4}). 

This research adopts the AB magnitude system \citep{Oke1983} and a \citet{Chabrier2003} stellar initial mass function (IMF). 
Also, we assume cosmological parameters of $\Omega_M=0.310$, $\Omega_\Lambda=0.689$, and $H_0=67.7$ km~s$^{-1}$Mpc$^{-1}$ in a flat Lambda cold dark matter model, which are consistent with those from the Planck 2018 VI results \citep{Planck2020}.

%%%%%%%%%%%%%%%%% DATA %%%%%%%%%%%%%%%%%%
\section{Data and methodology}\label{s2}

The source catalog and data set underlying this paper are available on the HSC-SSP Public Data Release site\footnote{\url{https://hsc.mtk.nao.ac.jp/ssp/data-release/}}. 
This section briefly overviews the narrow-band emitter catalog in HSC-SSP PDR2 \citep{Hayashi2020} and describes our procedure of image stacking of broad-band and narrow-band data for the targets, H$\alpha$ emitters at $z=0.4$. 

\subsection{HSC-SSP PDR2}\label{s21}

% figure 1
\begin{figure}
\begin{center}
\includegraphics[width=7.5cm]{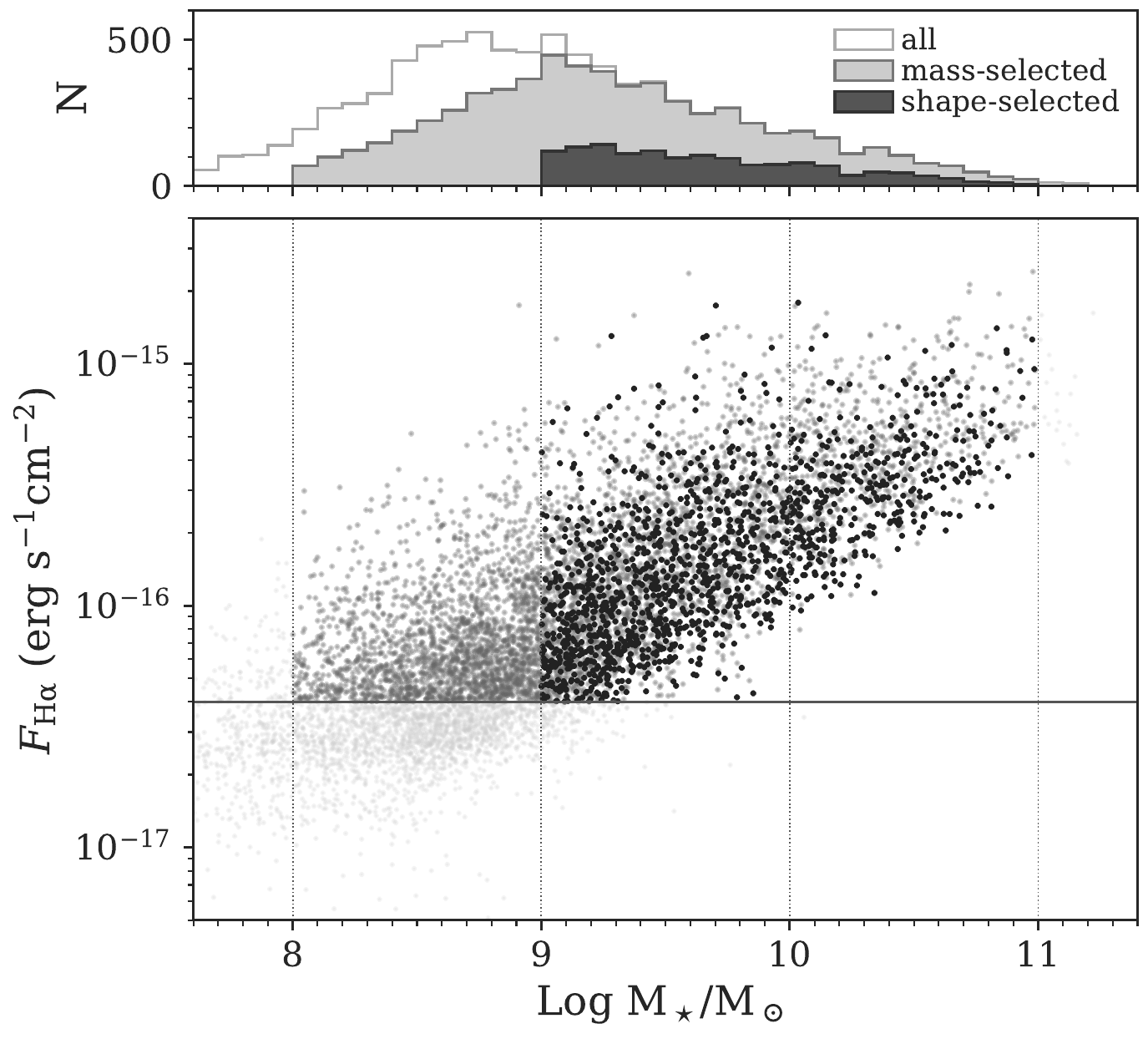}
\end{center}
\caption{
Narrow-band (H$\alpha$+[N{\sc ii}]) flux versus logarithmic stellar mass with stellar mass distributions on the top panel. 
The light-gray, gray, and black symbols indicate unused emitters in this work, those for the main analysis (section~\ref{s3}), and emitters with high ellipticities ($\epsilon>0.33$) that are used in the discussion section (see section~\ref{s4}). 
The horizontal line depicts the selection threshold for the narrow-band flux ($4\times10^{-17}$ erg~s$^{-1}$cm$^{-2}$). 
The dotted vertical lines represent selection cuts into three stellar-mass bins. 
}
\label{fig1}
\end{figure}

% table 1
\begin{table}
\caption{
Summary of the sample selection from the original source catalog (\cite{Hayashi2020}; $N=8625$).
See also figure~\ref{fig1} for the selection thresholds.}
\label{tab1}
\begin{center}
	\begin{tabular}{llr} % four columns, alignment for each
	\hline
	Sample        & Limits  & N\\
	\hline
	All       & $f_\mathrm{NB}$$>$4$\times$10$^{-17}$, FWHM$<$0.8 & 6324\\
	\hline
	Low-mass  & M$_\star$=$10^8$-$10^9$       & 2121\\
	Mid-mass  & M$_\star$=$10^9$-$10^{10}$    & 3139\\
	High-mass & M$_\star$=$10^{10}$-$10^{11}$ &  948\\
	Shape-aligned & M$_\star$=$10^9$-$10^{11}$, $\epsilon$$>$0.33 & 1430\\
	\hline
	\end{tabular}
\end{center}
\end{table}

This work is based on the H$\alpha$ emitter sample at $z=0.4$ established by \citet{Hayashi2020}; they reported narrow-band selected emission-line objects at $z<2$ based on the Deep and Ultradeep layers in HSC-SSP PDR2 \citep{Aihara2019}. 
The catalog contains 8625 H$\alpha$ emitters down to a limiting flux of $\sim1\times10^{-17}$ erg~s$^{-1}$cm$^{-2}$ (figure~\ref{fig1}) and line equivalent width (EW$_\mathrm{H\alpha+[N{\scriptscriptstyle II}]}$) of $>25$ \AA\ in the rest-frame \citep[tab.~4]{Hayashi2020}. 
They show flux excesses in narrow-band, NB921 ($\lambda_\mathrm{center}=9205$ \AA), relative to $z$ and $y$-bands by capturing H$\alpha$ ([N{\sc ii}] doublet also falls into the filter, however, we omit it hereafter) emission line at $z=$0.393--0.404 \citep[tab.~3]{Hayashi2020}. 
The survey area covers 22.09 deg$^2$ (16.79 deg$^2$ excluding bright star mask regions) split into 4 fields (Extended COSMOS, SXDS, ELAIS-N1, DEEP2-3; \cite[table~1]{Hayashi2020}). 
Stellar masses of our sample are derived by {\tt Mizuki}, a SED-based photo-$z$ code \citep{Tanaka2015,Tanaka2018}, by fixing the source redshift of $z=0.4$ ({\tt stellar\_mass\_mizuki\_zfixed\_convflux} in the source catalog; \cite{Hayashi2020}). 

To collect H$\alpha$ emitters homogeneously across the entire survey area, we set several selection criteria as follows. We adopt only H$\alpha$ emitters with narrow-band fluxes greater than $4\times10^{-17}$ erg~s$^{-1}$cm$^{-2}$ (SFR $=0.1$ M$_\odot$yr$^{-1}$ without dust correction), where the imaging depths achieve approximately 100 percent completeness over the whole survey fields \citep{Hayashi2020}. 
Also, this study does not consider areas taken under relatively bad seeing conditions (FWHM $>0.8$ arcsec) in either of narrow-band (NB921) or $z$-band. We then select 6324 H$\alpha$ emitters through these thresholds.

\subsection{Stacking analysis}\label{s22}

We perform median stacking for narrow-band (NB921) and $z$-band images of the selected H$\alpha$ emitters at $z=0.4$ to derive their typical radial profiles of H$\alpha$ and stellar surface densities. 
For the stacking analysis, we obtain coadd images in the NB921 and $z$-band of H$\alpha$ emitters from the Third Public Release of HSC-SSP \citep{Aihara2021}\footnote{\url{https://hsc-release.mtk.nao.ac.jp/das_cutout/pdr3/} (registration is required)}. 
The data were generated by the dedicated reduction pipeline ({\tt hscPipe} version 8; \cite{Bosch2018}). Their spatial resolutions are matched to FWHM $=0.8$ arcsec by Gaussian smoothing. 
We confirm that the composite PSF distributions of the NB921 and $z$-band are consistent within a margin of less than 6 percent, which is negligibly small for this study. 
Consequently, we derived median pixel values of the target images in each filter. 
Here, we separate the emitter sample into three stellar mass bins ($10^{8\mbox{-}9}$, $10^{9\mbox{-}10}$,$10^{10\mbox{-}11}$ M$_\odot$; see figure~\ref{fig1} and table~\ref{tab1}). 
Based on the composite NB921 and $z$ images, we establish emission line and stellar continuum maps through the following calculation. 
\begin{eqnarray}
F_\mathrm{H\alpha} &=& \Delta_\mathrm{NB} \frac{f_\mathrm{NB}-f_\mathrm{BB}}{1-\Delta_\mathrm{NB}/\Delta_\mathrm{BB}},\\
f_\mathrm{c} &=& \frac{f_\mathrm{BB}-f_\mathrm{NB}\cdot\Delta_{NB}/\Delta_\mathrm{BB}}{1-\Delta_\mathrm{NB}/\Delta_\mathrm{BB}},
\end{eqnarray}
where $f_\mathrm{NB}$ and $f_\mathrm{BB}$ are flux densities of the NB921 and $z$-band, and $\Delta_\mathrm{NB}$ and $\Delta_\mathrm{BB}$ indicate filter band widths of the NB921 and $z$-band, respectively \citep{Hayashi2020}. 
Furthermore, we adopt a single color-term correction of $f_{BB}=1.04\times f_{z}$ to correct a difference of the central filter wavelength between the NB921 and $z$-band (296 \AA), which is consistent with a typical value of high-mass H$\alpha$ emitters. 
Although the fixed color-term correction does not significantly affect our main result, this is an important issue on the discussion part regarding to the radial profile of EW$_\mathrm{H\alpha}$ (see section~\ref{s4} and Appendix~\ref{a1}). 

Besides, we generate random 100 realizations of composite emission-line and stellar continuum images to measure flux uncertainties of the stacked images using a bootstrap re-sampling from the same samples but with replacement. 
Based on the 100 random combined images, we calculate a standard deviation of self flux counts at a given radius. 
We also derive a background deviation given a pixel area ($S$) based on randomly-positioned empty apertures. 
Obtained background deviations correlate with the background areas by $\propto S^{\sim0.8}$, suggesting partial pixel-to-pixel correlations (cf. no correlation and full correlation if $\propto S^{0.5}$ and $\propto S$, respectively). 
This research defines the mean square errors of deviations of the self-flux and background noises as a total uncertainty of derived flux surface density (section~\ref{s3}). 
In addition, residual background counts are sampled by measuring median pixel values in the backgrounds of the randomly combined images. 
Obtained residual background counts are $\sim4\times10^{-4}$ and $\sim4\times10^{-3}$ in H$\alpha$ and continuum images, respectively, in the zero point magnitude of 27 mag. 
These residual backgrounds are subtracted when deriving the flux densities.

%%%%%%%%%%%%%%%%% RESULTS %%%%%%%%%%%%%%%%%%
\section{Results}\label{s3}

% figure 2
\begin{figure*}
\begin{center}
\includegraphics[width=16cm]{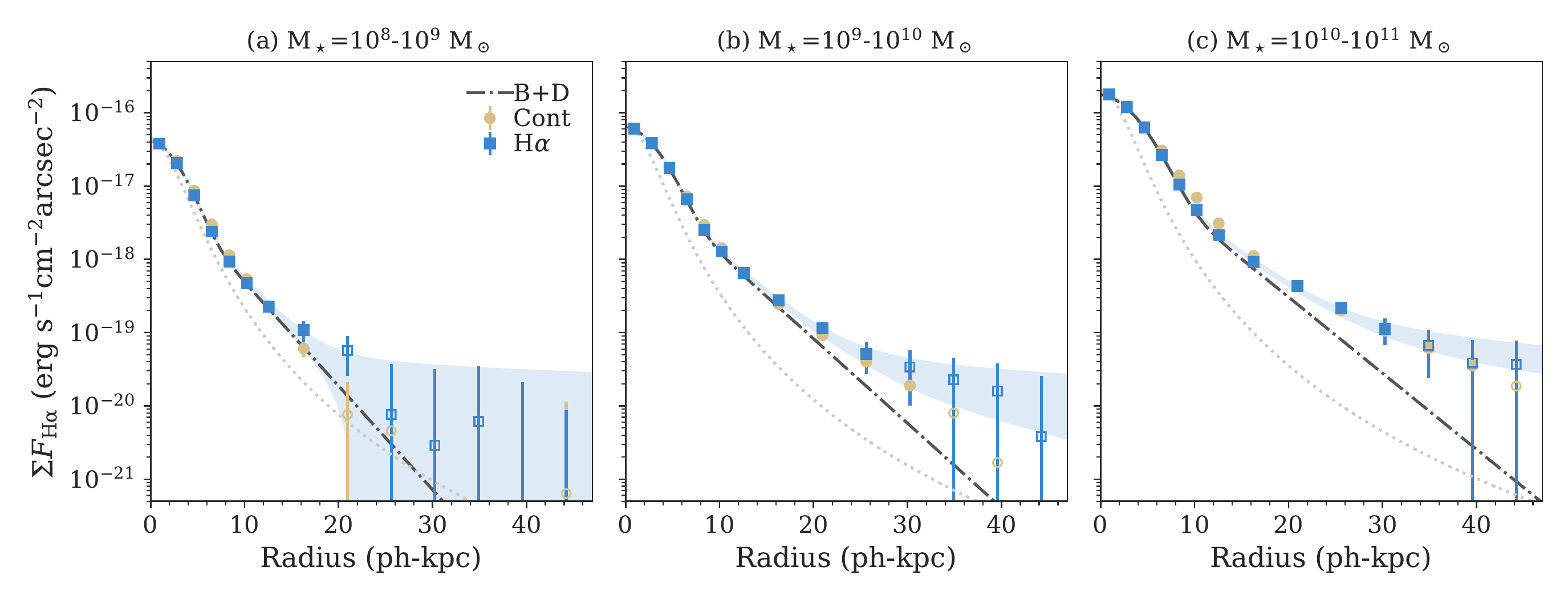}
\end{center}
\caption{
Radial profiles of surface brightness of H$\alpha$ line emission (blue squares) and normalized stellar continuum (yellow circles) for H$\alpha$ emitters at $z=0.4$. 
Less than $2\sigma$ detections are represented by the open symbols. 
The error-bars depict $1\sigma$ errors and the light-blue regions are $1\sigma$ errors of the curve fitting (see text). 
From the left to right, figures depict radial profiles for H$\alpha$ emitters with stellar masses of (a) $10^{8\mbox{-}9}$, (b) $10^{9\mbox{-}10}$, and (c) $10^{10\mbox{-}11}$ M$_\odot$, respectively. 
The dotted curves indicate scaled radial profiles of the composite PSF (fitted by a Moffat distribution). 
The dot-dashed curves are the best-fit Gaussian$+$exponential radial profiles, which are expected to trace bulge$+$disk components of the H$\alpha$ emitters. 
}
\label{fig2}
\end{figure*}

Median radial profiles of H$\alpha$ emission ($\Sigma F_\mathrm{H\alpha}$) and stellar continuum in three different stellar masses are shown in figure~\ref{fig2}, where radial continuum distributions are normalized at the peak surface densities in H$\alpha$ line. 
Examples of the composite H$\alpha$ and continuum images are represented in Appendix~\ref{a1}.
As a result, we detect both H$\alpha$ and continuum of the sample in outer regions of H$\alpha$ emitters at $z=0.4$, especially at the highest stellar mass bin ($10^{10\mbox{-}11}$ M$_\odot$). 
While we observe differentials between H$\alpha$ and stellar distributions in $\sim10$ ph-kpc at the highest stellar mass as explored in more detail in section~\ref{s4}, they are generally consistent with each other. 

To investigate stellar halo contributions, we fit the radial profiles with a combined function of Gaussian ($G$), exponential ($E$), and power law ($P$) distributions by assuming that they respectively trace galactic bulge, disk, and stellar halo: 
\begin{eqnarray}
G(r) &=& A_g \exp(-r^2 / w_g),\\
E(r) &=& A_e \exp(-r / r_s),\\
P(r) &=& A_p r^\gamma,
\end{eqnarray}
where $r$ is the physical radius (physical kpc, hereafter ph-kpc) from the center; $A_{g}$, $A_{e}$, $A_{p}$, $w_g$, $r_s$, and $\gamma$ are fitting parameters. 
We do not adopt the de Vaucouleurs-law profile \citep{Vaucouleurs1948} for bulge components because our targets are barely resolved or unresolved in the seeing-limited data (FWHM $=4.4$ ph-kpc). We perform a chi-square fitting with the model function shown above using a curve fitting code, {\tt lmfit} \citep{Newville2014}. 

% table 2
\begin{table}
\caption{
Best-fit parameters and $1\sigma$ errors in three different stellar mass bins 
(see figure~\ref{fig2}).}
\label{tab2}
\begin{center}
	\begin{tabular}{lrrr} % four columns, alignment for each
	\hline
	Mass range  & $10^8$--$10^9$ & $10^9$--$10^{10}$ & $10^{10}$--$10^{11}$\\
	\multicolumn{1}{c}{($N=$)} & (2121) & (3139) & (948)\\
	\hline
	$A_{g}$ ($\times10^{-17}$) & $3.0\pm0.2$  & $4.9\pm0.2$  & $14.2\pm0.6$\\
	$w_{g}$                    & $11.6\pm0.5$ & $15.9\pm0.5$ & $20.9\pm0.8$\\
	$A_{e}$ ($\times10^{-17}$) & $1.3\pm0.4$  & $1.7\pm0.4$  & $3.7\pm1.0$\\
	$r_{s}$                    & $3.1\pm0.3$  & $3.8\pm0.3$  & $4.2\pm0.4$\\
	$A_{p}$ ($\times10^{-18}$) & $0.6\pm2.1$  & $1.5\pm2.6$  & $11.8\pm9.1$\\
	$\gamma$                   & $-1.2\pm1.3$ & $-1.2\pm0.5$ & $-1.4\pm0.2$\\
	\hline
	\end{tabular}
\end{center}
\end{table}

The best-fit parameters and associated errors for the H$\alpha$ flux surface densities are summarized in table~\ref{tab2}. 
Figure~\ref{fig2} shows that there are residual components beyond $\gtrsim15$ ph-kpc compared to the best-fit bulge$+$disk profiles in the intermediate and high stellar mass bins (figure~\ref{fig2}b,c). 
Such residuals have been commonly observed in stellar distributions of nearby galaxies (e.g., \cite{Courteau2011,Merritt2016}), suggesting that they could originate respectively from in-situ hot young stars and old stars in outer disks and stellar halos (see section~\ref{s4} for more discussion). 
The estimated power law slopes in the outskirts, $\gamma=-1.4\sim-1.2$, are flatter than those of most nearby galaxies $\gamma<-2$ \citep{Harmsen2017}. 
However, given the low signal detection, it may be too early to conclude that H$\alpha$ emitters at $z=0.4$ have much shallower power-law slopes than galaxies in the local universe. 
Recent studies have reported that stellar halo distributions can be fitted by broken power law; the outer slopes become significantly steeper at broken radii of a few tens of ph-kpc \citep{Deason2014}. 
This may be one of the reasons of non detection in both H$\alpha$ and continuum images at $r>30$ ph-kpc. 

\begin{figure}
\begin{center}
\includegraphics[width=7.5cm]{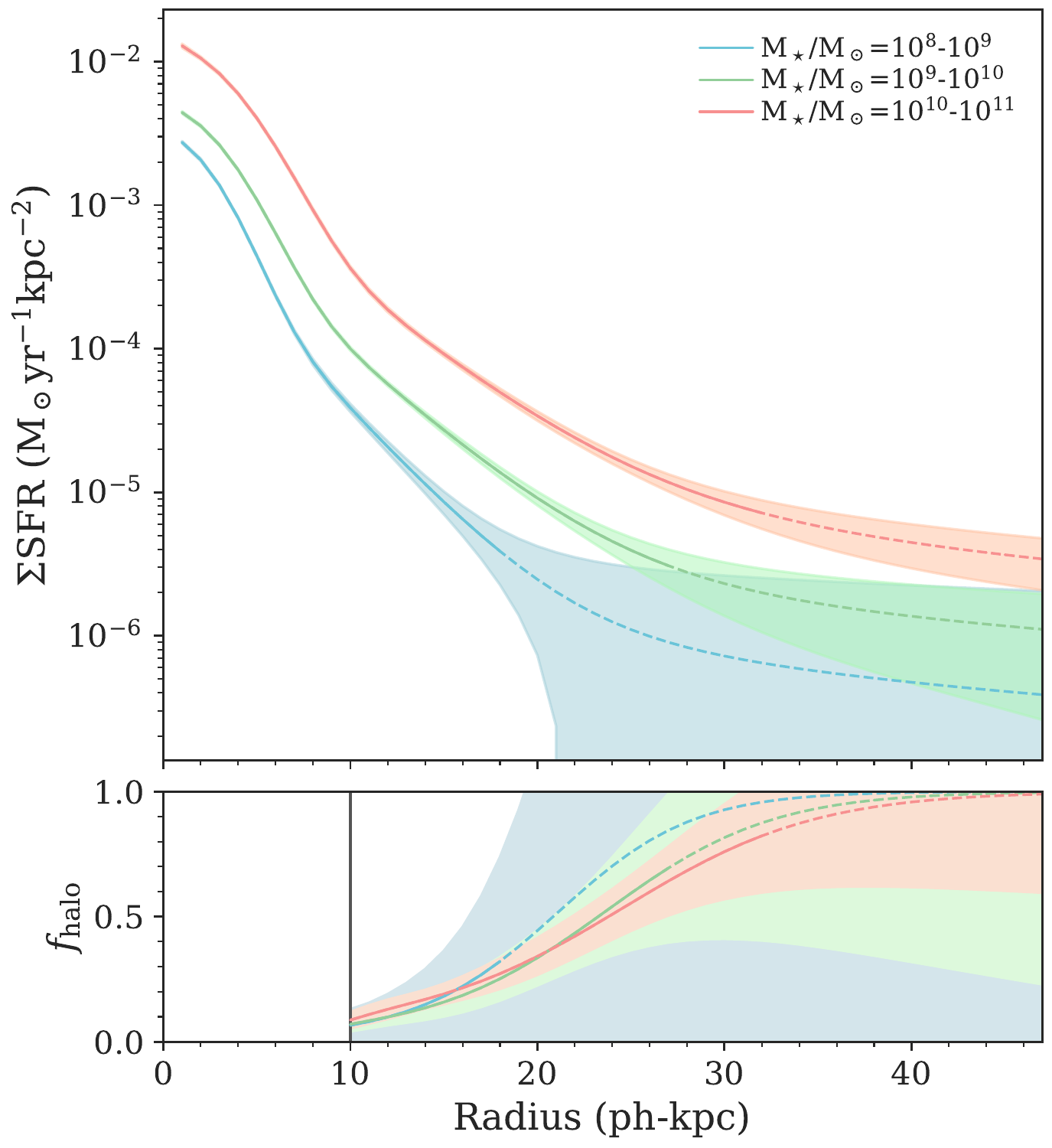}
\end{center}
\caption{
(Top panel) Radial profiles of surface SFR densities of H$\alpha$ emitters in three stellar mass bins as in figure~\ref{fig1} and table~\ref{tab1} (blue: $10^{8\mbox{-}9}$, green $10^{9\mbox{-}10}$, and red: $10^{10\mbox{-}11}$ M$_\odot$, respectively). 
(Bottom panel) Fractions of power law components to the best-fitted radial profiles ($f_\mathrm{halo}$) in respective stellar mass bins at $r>10$ ph-kpc. 
Non detection regions with $<2\sigma$ are extrapolated by the best-fitted curves (dashed curves). 
}
\label{fig3}
\end{figure}

Moreover, we convert the H$\alpha$ surface densities to SFR surface densities ($\Sigma\mathrm{SFR}$) and investigate fractions of star formation in stellar halos at $>10$ ph-kpc (figure~\ref{fig3}). 
We follow the \citet{Kennicutt1998} calibration and scale by $1/1.7$ to apply the \citet{Chabrier2003} IMF. 
Also, we assume 20 percent contributions of [N{\sc ii}]$\lambda\lambda6548,6583$ lines to narrow-band fluxes. 
Here it should be noted that we assume no extinction and do not consider warm ionized medium \citep{Reynolds1973,Haffner2009}. 
Thus, the derived $\Sigma\mathrm{SFR}$ values should have significant uncertainties beyond the measurement errors. 
In particular, H$\alpha$ extinction in the central components of emitters would exceed 1 mag at the high stellar mass bin \citep{Sobral2016}. 
However, central star formation is not the scope of this paper; therefore, we solely focus on the outskirts of H$\alpha$ emitters. 
The stacked data have reached $\Sigma\mathrm{SFR}$ down to $\sim3\times10^{-6}$ M$_\odot$yr$^{-1}$kpc$^{-2}$, about a half of which appears to originate from stellar halos. 
As inferred from the best-fit results, major fluxes beyond 30 ph-kpc are dominated by halo components; however, significant uncertainties are associated with these results. 

% figure 3

%%%%%%%%%%%%%%%%% DISCUSSION %%%%%%%%%%%%%%%%%%
\section{Discussion and summary}\label{s4}

This work has identified diffuse H$\alpha$ emission and stellar continuum beyond 10 ph-kpc and up to 30 ph-kpc for the H$\alpha$ emitters at $z=0.4$ based on the deep narrow-band and broad-band stacking taken from HSC-SSP. 
There are residual excesses in both H$\alpha$ line and stellar continuum compared to exponential distributions at $\gtrsim20$ ph-kpc. 
While such residual emissions could originate respectively from in-situ hot young stars and old stars in outer disks and stellar halos, we cannot ignore contributions from associated halos \citep{Zhang2016,Zhang2018} and escape of ionizing radiation (e.g., \cite{Leitherer1995,Heckman2011}).
Thus physical origins of diffuse emission lines are still controversial. 
In addition, we do not obtain a clear discrepancy between the radial profiles of the emission line and the continuum on the outskirts. 
The observed H$\alpha$ surface densities monotonically decline in the outskirts with flatter power-law slopes of $\gamma=-1.4\sim-1.2$ than those in nearby galaxies; however, we note weak detection of some diffuse stellar halos. 

Lastly, we delve into causal factors that may contribute to the observed diffuse H$\alpha$ emission on the outskirt by constructing shape-aligned composite images of H$\alpha$ emitters. 
This is originally motivated to observe diffuse ionized gas arisen from biconical outflows from galaxy disks as typified by nearby starburst, M82 (e.g., \cite{Lynds1963,Visvanathan1972,Bland1988,Veilleux2003,Yoshida2019}). 
Based on the second moments of the object intensity termed adaptive moments, we calculate ellipticities ($\epsilon$) and position angles of the targets \citep{Bernstein2002}. 
We use the adaptive moments in the $i$-band ({\tt i\_sdssshape\_shape}) because the $i$-band data have better and more homogeneous seeing sizes ($\sim0.6$ arcsec) over the survey field. 
Also, we employ only elliptic sources at M$_\star>10^{9}$ M$_\odot$ with $\epsilon>0.33$ (i.e., major-axis $>1.5\times$ minor-axis) to assure the credibility of position angle estimates. 
We align objects along the major-axis; consequently, we conduct the median stacking following a similar method described in section~\ref{s22} for 1430 objects that satisfy these selection criteria (see figure~\ref{fig1} and table~\ref{tab1}). 

\begin{figure}
\begin{center}
\includegraphics[width=7.5cm]{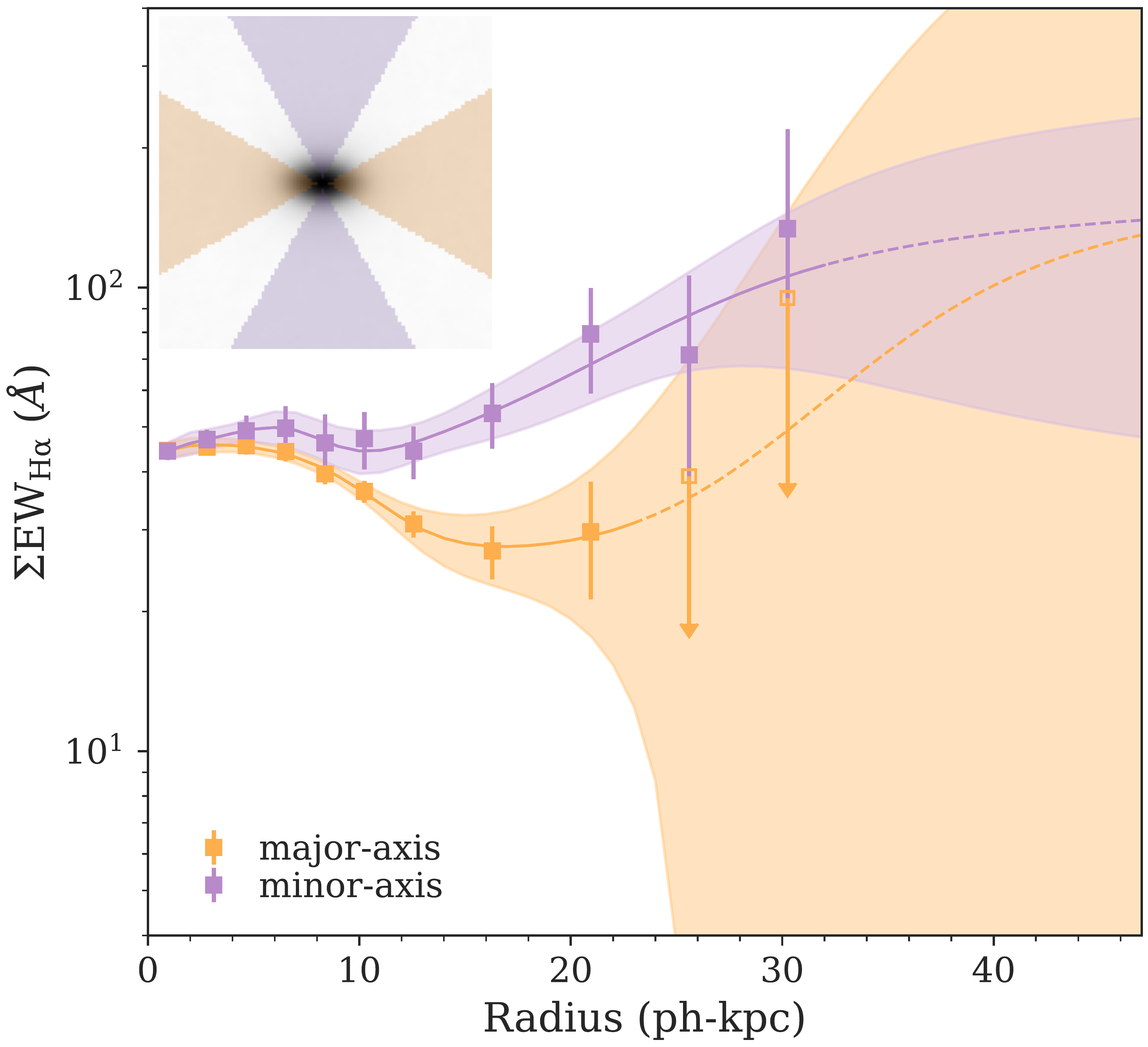}
\end{center}
\caption{
Radial profiles of rest-frame EW of H$\alpha$ emission in the directions of $\pm30$ deg from major-axis (orange symbols) and minor-axis (purple symbols), respectively. 
An image illustration of the areas used in the stacking analysis is denoted at the upper left corner. The error-bars and color-filled regions represent $1\sigma$ errors. 
The open symbols indicate $2\sigma$ upper limits. 
The dashed curves depict extrapolated lines of the best-fitted models. 
The gray region indicates the standard deviations of non-aligned EW$_\mathrm{H\alpha}$ profile based on 100 composite images with random rotation. 
}
\label{fig4}
\end{figure}

Figure~\ref{fig4} represents radial profiles of the rest-frame EW$_\mathrm{H\alpha}$ along $\pm30$ deg of the major- and minor-axis. 
For deriving radial EW$_\mathrm{H\alpha}$ profiles appropriately, the color term correction is important because the color term is significantly different between the major- and minor-axis. 
We thus correct the color term effect in each radial bin on the best effort basis, by using the control sample (see Appendix~\ref{a1} for the detailed procedure). 
The result shows a clear directivity of EW$_\mathrm{H\alpha}$; for instance, EW$_\mathrm{H\alpha}$ is higher on the outskirts along the minor-axis of the stellar disk (figure~\ref{fig4}). 
A possible factor producing such high EW$_\mathrm{H\alpha}$ could be biconical galactic outflows. 
However, as discussed above, there are various potential contributions such as star formation in outer disks and emission from associated halos; therefore, the actual cause would be more complex and complicated. 
On the other hand, the moderate dip in the center can be explained by higher nebular dust extinction and/or lower specific star formation rate in the galactic bulge. 
%In addition, to inspect a systematic effect, we estimate standard deviations of radial EW$_\mathrm{H\alpha}$ distributions of composite images with random rotation (0--90 deg). 

In summary, we have unveiled stellar and H$\alpha$ radial profiles on the outskirts of H$\alpha$ emitters at $z=0.4$ for the first time. 
However, contributors to diffuse H$\alpha$ emission remain uncertain. 
Comparing our results with hydrodynamic simulations and absorption-line observations (e.g., \cite{Werk2014,Prochaska2017}) will help us resolve its causal factors from theoretical and independent points of view, respectively. 
In addition, more detailed investigations of physical properties of diffuse halos, e.g., derivation of stellar populations on the outskirts based on multi-band image stacking, will deliver further insights into mass-assembly histories surrounding galaxies beyond the local universe.

% figure 4

%%%%%%%%%%%%%%%%% ACKNOWLEDGEMENTS %%%%%%%%%%%%%%%%%%

\begin{ack}
Based on data collected at the Subaru Telescope and retrieved from the HSC data archive system, which is operated by Subaru Telescope and Astronomy Data Center at National Astronomical Observatory of Japan. We are honored and grateful for the opportunity of observing the Universe from Maunakea, which has the cultural, historical and natural significance in Hawaii. 

The Hyper Suprime-Cam (HSC) collaboration includes the astronomical communities of Japan and Taiwan, and Princeton University. The HSC instrumentation and software were developed by the National Astronomical Observatory of Japan (NAOJ), the Kavli Institute for the Physics and Mathematics of the Universe (Kavli IPMU), the University of Tokyo, the High Energy Accelerator Research Organization (KEK), the Academia Sinica Institute for Astronomy and Astrophysics in Taiwan (ASIAA), and Princeton University. Funding was contributed by the FIRST program from Japanese Cabinet Office, the Ministry of Education, Culture, Sports, Science and Technology (MEXT), the Japan Society for the Promotion of Science (JSPS), Japan Science and Technology Agency (JST), the Toray Science Foundation, NAOJ, Kavli IPMU, KEK, ASIAA, and Princeton University. 
This paper makes use of software developed for the Large Synoptic Survey Telescope. We thank the LSST Project for making their code available as free software at \url{http://dm.lsst.org}

The Pan-STARRS1 Surveys (PS1) have been made possible through contributions of the Institute for Astronomy, the University of Hawaii, the Pan-STARRS Project Office, the Max-Planck Society and its participating institutes, the Max Planck Institute for Astronomy, Heidelberg and the Max Planck Institute for Extraterrestrial Physics, Garching, The Johns Hopkins University, Durham University, the University of Edinburgh, Queen’s University Belfast, the Harvard-Smithsonian Center for Astrophysics, the Las Cumbres Observatory Global Telescope Network Incorporated, the National Central University of Taiwan, the Space Telescope Science Institute, the National Aeronautics and Space Administration under Grant No. NNX08AR22G issued through the Planetary Science Division of the NASA Science Mission Directorate, the National Science Foundation under Grant No. AST-1238877, the University of Maryland, and Eotvos Lorand University (ELTE) and the Los Alamos National Laboratory.

%We thank anonymous referee for helpful feedback.

This research was financially supported by JSPS KAKENHI Grant Number JP19K14766. 
We would like to thank Editage (\url{www.editage.com}) for English language editing. 
This work made extensive use of the following tools, {\tt NumPy} 
\citep{Harris2020}, {\tt Matplotlib} \citep{Hunter2007}, {\tt lmfit} 
\citep{Newville2014}, the Tool for OPerations on Catalogues And Tables, 
{\tt TOPCAT} \citep{Taylor2005}, a community-developed core Python package for 
Astronomy, {\tt Astopy} \citep{Astropy2013}, and Python Data Analysis Library 
{\tt pandas} \citep{McKinney2010}. 
\end{ack}

%%%%%%%%%%%%%%%%% APPENDIX %%%%%%%%%%%%%%%%%%
\appendix 

\section{Radial color term dependence}\label{a1}

\begin{figure}
\begin{center}
\includegraphics[width=7.5cm]{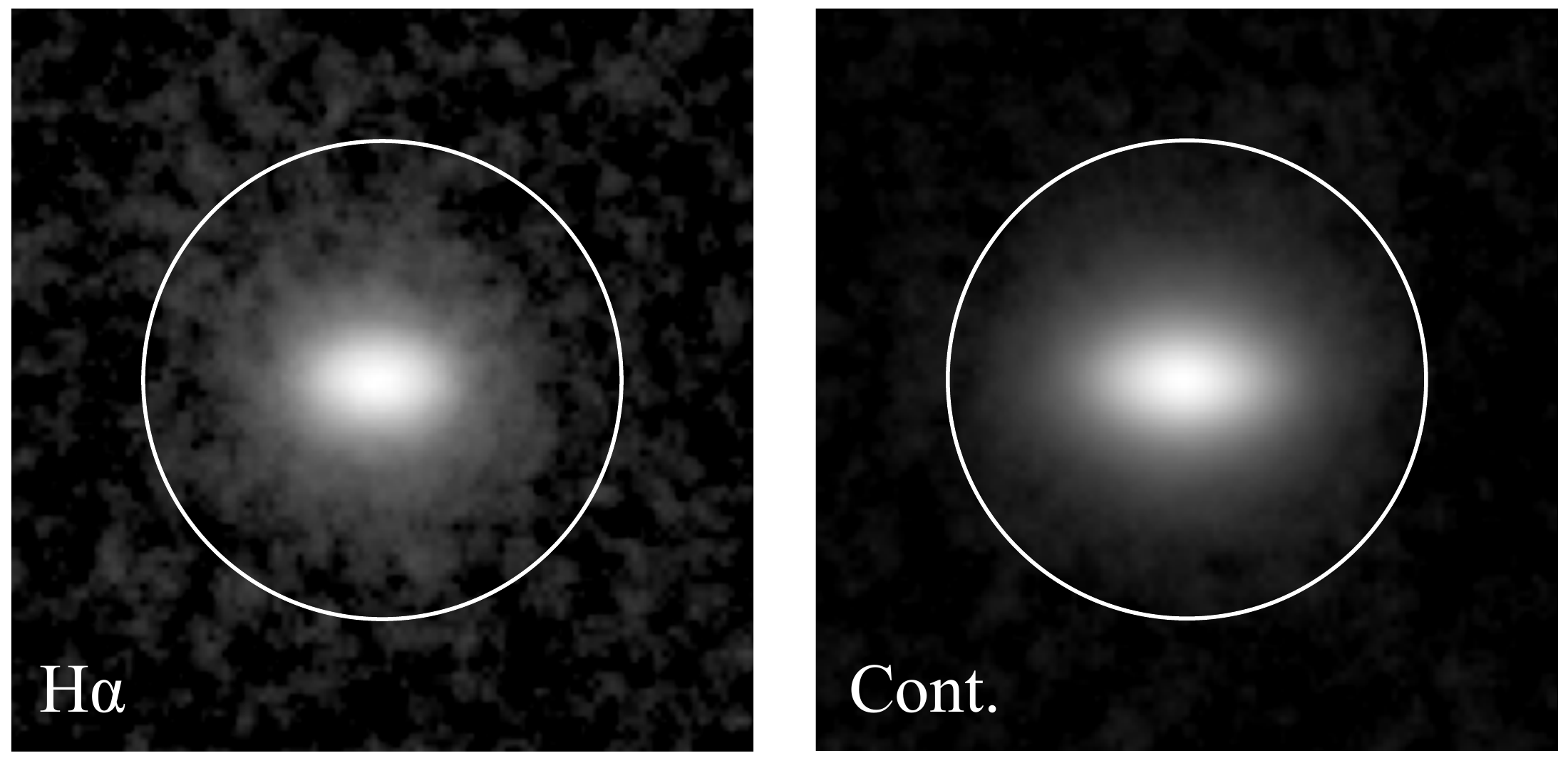}
\end{center}
\caption{
The left and right panels show the composite H$\alpha$ (+[N{\sc ii}]) and continuum images of shape-aligned H$\alpha$ emitters (see section~\ref{s4}). 
The white circles depict the radius of 30 ph-kpc. 
Here they are stretched to a logarithmic scale to improve visibility. 
}
\label{fig1a}
\end{figure}

This section describes how we determine the color term values between NB921 and $z$-band to investigate the H$\alpha$ directionality for our H$\alpha$ emitters at $z=0.4$ in section~\ref{s4}. 
The shape-aligned composite H$\alpha$ and continuum images of H$\alpha$ emitters are shown in figure~\ref{fig1a}, where we assume a fixed color term value of $\zeta\equiv f_\mathrm{NB}/f_{z}=1.0375$. 
The original emitter source catalog \citep{Hayashi2020} applied the color-term correction by using weighted combined $zy$ fluxes, in particular, to separate [O{\sc ii}] emitters at $z=1.47$ from distant red galaxies at $z\sim1.3$. 
However, we decided not to use $y$-band data in the stacking analyses taking into account its by $\sim1$ mag shallower imaging depth compared to that in the $z$-band. 
In addition, we detect significantly brighter PSF wings in the $y$-band than those in the NB921 and $z$-bands, which make the issue even more complicated.

% figure 1a

\begin{figure}
\begin{center}
\includegraphics[width=7.5cm]{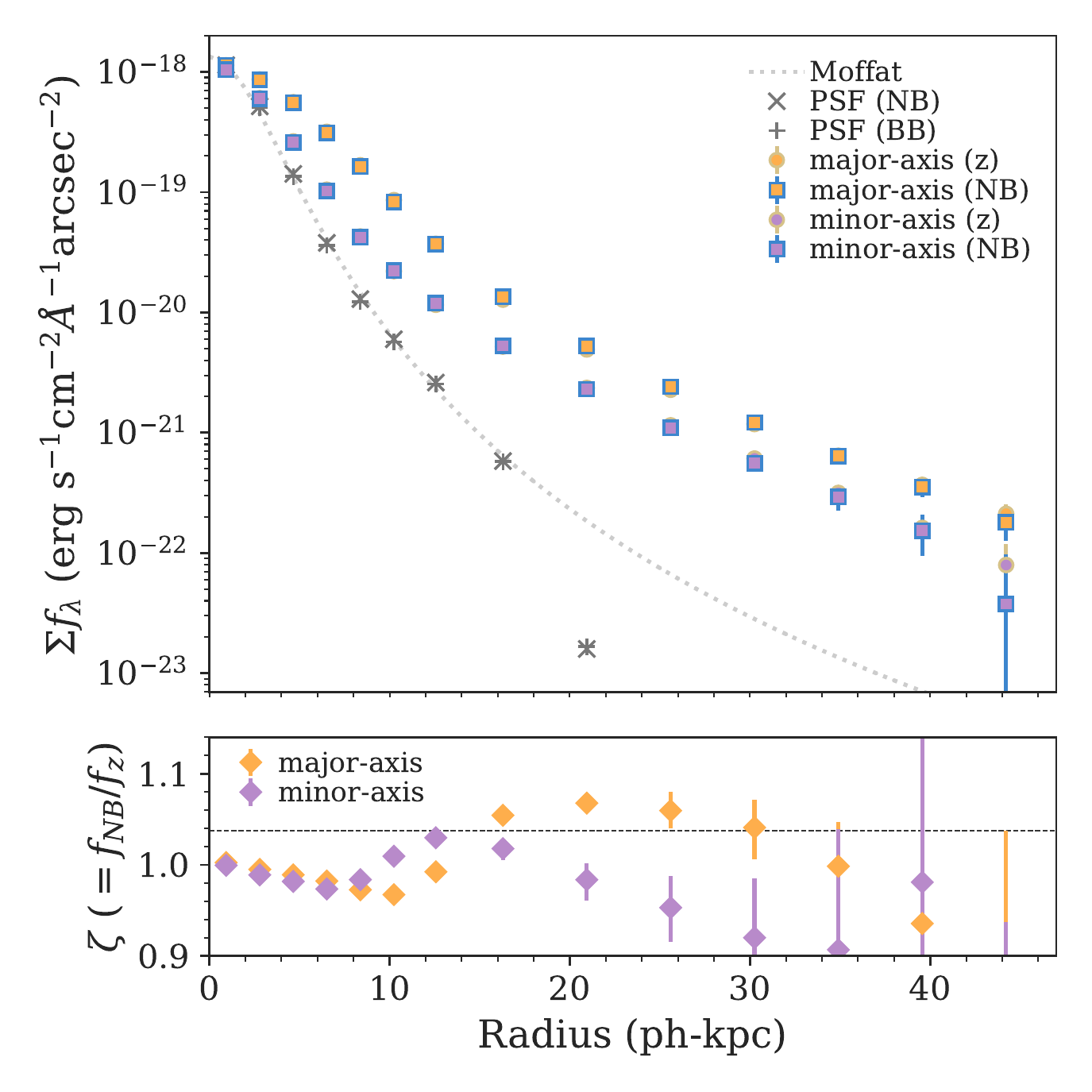}
\end{center}
\caption{
(Top panel) Radial profiles of surface flux densities in NB921 ($f_\mathrm{NB}$) and $z$-band ($f_\mathrm{z}$). 
The orange squares and circles respectively indicate $f_\mathrm{NB}$ and $f_\mathrm{z}$ in the major-axis. 
Similarly, the purple squares and circles show $f_\mathrm{NB}$ and $f_\mathrm{z}$ in the minor-axis. 
The black cross and plus symbols depict composite PSFs in NB921 and $z$-band, respectively. 
The dotted line is the best-fit Moffat model to the composite PSF in NB921, adopted throughout the paper. 
(Bottom panel) Radial color term distribution along the major-axis (orange) and minor-axis (purple). 
In both panels, error-bars indicate 68 percentiles in 100 random realizations. 
}
\label{fig2a}
\end{figure}

Instead, we derive typical color term values by using 37163 non-emitter samples with similar stellar populations (M$_\star=10^{9-11}$ M$_\odot$ and SFR $>0.1$ M$_\odot$yr$^{-1}$) at $z_\mathrm{photo}=$ 0.3--0.5 in the same field. 
Their stellar masses, SFRs, and photometric redshifts are obtained from {\tt Mizuki} SED fitting with reduced chi-squares $<5$ \citep{Tanaka2015,Tanaka2018}. 
The control samples also satisfy the same selection threshold of the shape measurement as for the shape-aligned H$\alpha$ emitter sample ($\epsilon>0.33$). 
We then generate 100 realizations of shape-aligned stacked NB921 and $z$ images of randomly selected 10000 non-emitters from the control samples. 
We here adjust the sampling rates of non-emitters to match the stellar mass distribution to the H$\alpha$ emitter sample. 
Based on random composite images, we measure the median color term values, $\zeta(r)\equiv f_\mathrm{NB}(r)/f_{z}(r)$, in each radial bin along the major- and minor-axis (figure~\ref{fig2a}). 

\begin{figure*}
\begin{center}
\includegraphics[width=16cm]{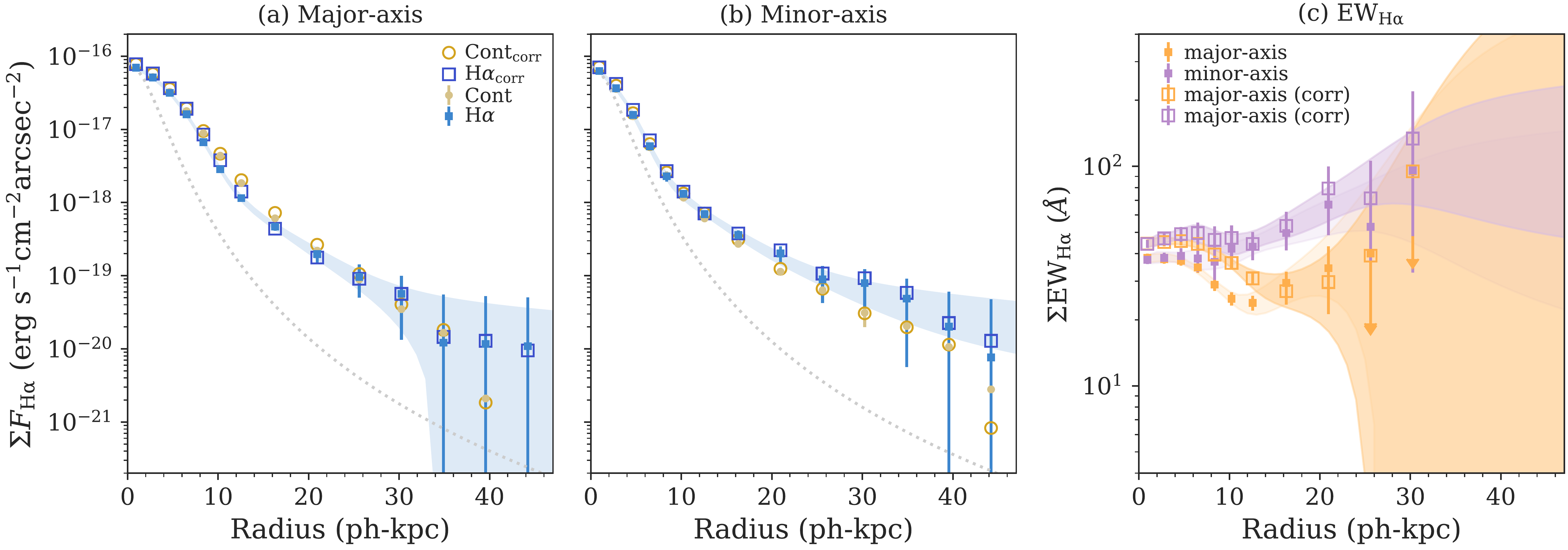}
\end{center}
\caption{
(a) Same as in figure~\ref{fig2} but for the major-axis of the shape-aligned composite images. 
Those with color-term corrections based on $\zeta(r)$ derived in figure~\ref{fig2a}, are shown by blue and yellow open symbols, respectively. 
(b) Same as in figure~\ref{fig3a}a, but for the minor-axis. 
(c) Same as in figure~\ref{fig4}, but we here compare radial profiles of EW$_\mathrm{H\alpha}$ with and without color term corrections, which are respectively depicted by open symbols with dark-colored regions and filled symbols with light-colored regions. 
}
\label{fig3a}
\end{figure*}

We adopt measured $\zeta(r)$ values to obtain color-term corrected radial profiles of EW$_\mathrm{H\alpha}$ in the major- and minor-axis (figure~\ref{fig3a}). 
First of all, figure~\ref{fig2a} indicates bluer color terms along the major-axis than the minor-axis at $\sim10$ ph-kpc, which should be caused by disk components of star-forming galaxies. 
Such a low color term leads to a strange dip of EW$_\mathrm{H\alpha}$ in the major-axis if we do not consider this effect (figure~\ref{fig3a}c). 
The color terms then increase at larger radii where old stellar populations are more dominant. 
In contrast, they tend to decline at $\gtrsim14$ ph-kpc in the minor-axis and $\gtrsim20$ ph-kpc in the major-axis. 
Such depressions are thought to be simply due to lower signals in NB921. 
However, one should note that the color term uncertainties in the faint end do not affect our conclusions given large EW$_\mathrm{H\alpha}$ errors (figure~\ref{fig3a}).

% figure 2a

% figure 3a

\bigskip

%%%
%%%%%%%%%%%%%%%%%%%% REFERENCES %%%%%%%%%%%%%%%%%%

\end{document}